\documentclass[twocolumn,showpacs,floatfix]{revtex4}%
\usepackage[dvipdfmx]{graphicx, color}
\usepackage{amsmath}%
\usepackage{amsfonts}%
\usepackage{amssymb}
\usepackage{bm}
\usepackage{here}

\def\0{{ {\bm 0} }}

\allowdisplaybreaks[4]

\begin{document}
\title{
Pairing mechanism for nodal $s$-wave superconductivity in BaFe$_2$(As,P)$_2$:
Analysis beyond Migdal-Eliashberg formalism
}
\author{
Hironori \textsc{Nakaoka}$^{1}$, 
Youichi \textsc{Yamakawa}$^{1}$, and 
Hiroshi \textsc{Kontani}$^{1}$
}

\date{\today }

\begin{abstract}
The pairing mechanism and gap structure in Ba122 pnictides
have been hotly discussed for long time
as one of the central issues in Fe-based superconductors.
Here, we attack this problem by taking account of the 
vertex corrections (VCs) for the Coulomb interaction $U$ ($U$-VCs), 
which are totally dropped in conventional Migdal-Eliashberg formalism.
The $U$-VC in the charge susceptibility 
induces strong orbital fluctuations, and the $U$-VC also 
enlarges the orbital-fluctuation-driven attractive interaction.
By analyzing the effective multiorbital Hubbard model for Ba122 pnictides, 
we find that the orbital fluctuations develop
in all four $d$-orbitals ($t_{2g}$- and $z^2$-orbitals), 
by which the FSs are composed.
For this reason, nearly isotropic gap function appears 
on all the hole-type FSs, 
including the outer hole-FS around Z-point composed of $z^2$-orbital.
In contrast, nodal gap structure appears on the electron-FSs
for wide parameter range.
The obtained nodal $s$-wave state changes to fully-gapped 
$s$-wave state without sign-reversal ($s_{++}$-wave state) 
by introducing small amount of impurities,
accompanied by small reduction in $T_{\rm c}$.
The present microscopic theory naturally explains the
important characteristics of the gap structure of both hole- and electron-FSs
in Ba122 pnictides, without introducing any phenomenological pairing interaction.
\end{abstract}

\address{
$^1$ Department of Physics, Nagoya University,
Furo-cho, Nagoya 464-8602, Japan. 
}
 
\pacs{71.45.Lr,74.25.Dw,74.70.-b}

\sloppy

\maketitle

\section{Introduction}
\label{sec:intro}
Ba122 pnictides have been studied for a long time
as typical Fe-based superconductors, which are strongly correlated multiorbital superconductors.
As possible pairing states,
both the spin-fluctuation-mediated $s_\pm$-wave state 
\cite{Kuroki2008,Mazin2008,Chubukov,Hirschfeld}
and the orbital-fluctuation-mediated $s_{++}$-wave state 
\cite{Kontani2010}
have been discussed in various Fe-based superconductors.
In many compounds, the hole- and electron-Fermi surfaces (FSs) 
are composed of only three $t_{2g}$-orbitals;
$xz$-, $yz$-, and $xy$-orbitals.
In the case of Ba122 compounds,
the $z^2$-orbital contributes to the outer cylinder hole-FS 
around Z point, in addition to $t_{2g}$-orbitals.
There is no $z^2$-orbital weight on electron-FSs.
The superconducting gap function on the $z^2$-orbital outer hole-FS
has been analyzed for years, as a key to understand the pairing mechanism \cite{Suzuki-Kuroki,Shimojima,Yoshida,Feng,Saito-loopnode}.

When the random-phase-approximation (RPA) was applied to 
the three-dimensional Ba122 system
\cite{Suzuki-Kuroki},
spin fluctuations develop only in the $t_{2g}$-orbitals,
whereas spin fluctuations in the $z^2$-orbital remain very small
due to the absence of inter-pocket nesting.
For this reason, in the obtained $s_\pm$-wave state,
the gap function of $z^2$-orbital outer hole-FS, 
$\Delta_{h,z^2}$, is small.
Thus, the horizontal node appears robustly within the RPA.
However, several ARPES studies
\cite{Shimojima,Yoshida}
reported the absence of the horizontal node,
that is, the relation $\Delta_{h,z^2} \sim \Delta_{h,t_{2g}}$ holds.
In contrast, presence of the horizontal node
($\Delta_{h,z^2}\cdot\Delta_{h,t_{2g}}<0$ and
$|\Delta_{h,z^2}| \ll \Delta_{h,t_{2g}}$)
was reported in Ref. \cite{Feng}.
To understand the relation 
$\Delta_{h,z^2} \sim \Delta_{h,t_{2g}}$,
the present authors showed that 
nearly orbital-independent fully-gapped $s_{++}$-wave state
is realized on all hole-FSs
when strong inter-orbital-fluctuations involving four $d$-orbitals emerge
 \cite{Saito-loopnode}.
Thus, presence or absence of the horizontal node in Ba122
is a significant key factor to distinguish the pairing mechanism.

To clarify the pairing mechanism,
it is significant to understand 
the origin of the electronic nematic order at $T_{S}$,
which is above the magnetic order temperature $T_N$.
Both the spin-nematic scenario
\cite{Kivelson,Fernandes} 
and the orbital order scenario
\cite{Kruger-OO,Lee-OO,Onari-SCVC,Kontani-Raman} 
have been discussed very actively.
In the latter scenario,
higher-order electronic correlations,
called the vertex corrections (VCs), 
should be taken into account.
In Ref. \cite{Onari-SCVC}, the authors found the 
Aslamazov-Larkin (AL) type VC for the bare Coulomb 
interaction ${\hat U}$, which we call the $U$-VC, 
induces the orbital order under moderate spin fluctuations
\cite{Onari-SCVC,Yamakawa-FeSe,Onari-FeSe}.
In Refs. \cite{Yamakawa-CDW,Tsuchiizu2016-cuprate},
this mechanism has been applied to explain the
nematic charge-density-wave in cuprate superconductors
\cite{Sachdev-CDW, Wang-CDW}.

In our previous study for Ba122 systems \cite{Saito-loopnode},
the $U$-VC had been neglected.
However, the $U$-VC due to the AL-VC is significant
not only for the charge susceptibility, 
but also for the electron-boson coupling in the gap equation.
In fact, the Migdal theorem cannot be applied to 
strongly-correlated superconductors with strong
spin/charge fluctuations.
In Refs. \cite{Onari-PRL2014,Tazai-PRB2016,Tazak-JPSJ2017,Yamakawa-PRB2017},
we have shown that the orbital-fluctuation-driven attractive
interaction is strongly enlarged by the AL-type $U$-VC.
In Fe-based superconductors, both
ferro- and antiferro-orbital fluctuations develop,
and the attractive pairing interaction is strongly 
magnified by the $U$-VC that is neglected in the Migdal approximation.
\cite{Onari-PRL2014,Yamakawa-PRB2017}.
In both La1111 and FeSe,
the $s_{++}$-wave state is naturally obtained
by introducing the $U$-VC into the gap equation,
by formulating the gap equation
going beyond the Migdal-Eliashberg (ME) formalism.

In Ref. \cite{Saito-loopnode},
we introduced a phenomenological inter-orbital quadrupole interaction
to realize the strong orbital fluctuations in four
$t_{2g}$- and $z^2$-orbitals
within the RPA.
However, it is highly nontrivial whether strong
orbital fluctuations appear in four $d$-orbitals comparably by including
the $U$-VC or not, based on the Ba122 model 
with on-site Hubbard interaction.
It is highly nontrivial theoretical challenge to 
explain the relation $\Delta_{h,z^2} \approx \Delta_{h,t_{2g}}$
based on the realistic Hubbard model for Ba122 systems.

In the present paper,
we revisit the study of pairing mechanism and gap structure in
BaFe$_2$(As,P)$_2$,
which has been discussed for years
as one of the central issue in Fe-based superconductors.
For this purpose, 
we construct the effective two-dimensional tight-binding model 
for BaFe$_2$(As,P)$_2$, in which the FSs are composed of four
$t_{2g}$- and $z^2$-orbitals.
Based on this model, we analyze the electronic states based on 
the self-consistent vertex correction (SC-VC) method
\cite{Onari-SCVC}.
Due to the AL-type $U$-VC, strong ferro-orbital and
antiferro-orbital fluctuations emerge in four $d$-orbitals comparably.
For this reason, nearly isotropic gap function appears 
on all hole-FSs, including the $z^2$-orbital outer hole-FS.
In contrast, loop-nodes are expected to appear on the electron-FSs
for wide parameter range.
Thus, the present study satisfactorily explains 
two characteristics of the gap structure in BaFe$_2$(As,P)$_2$;
absence of horizontal node on hole-FSs and 
the presence of loop nodes on electron-FSs.

\section{model Hamiltonian}
\label{sec:model}
\begin{figure}[htbp]
\includegraphics[width=0.95\linewidth]{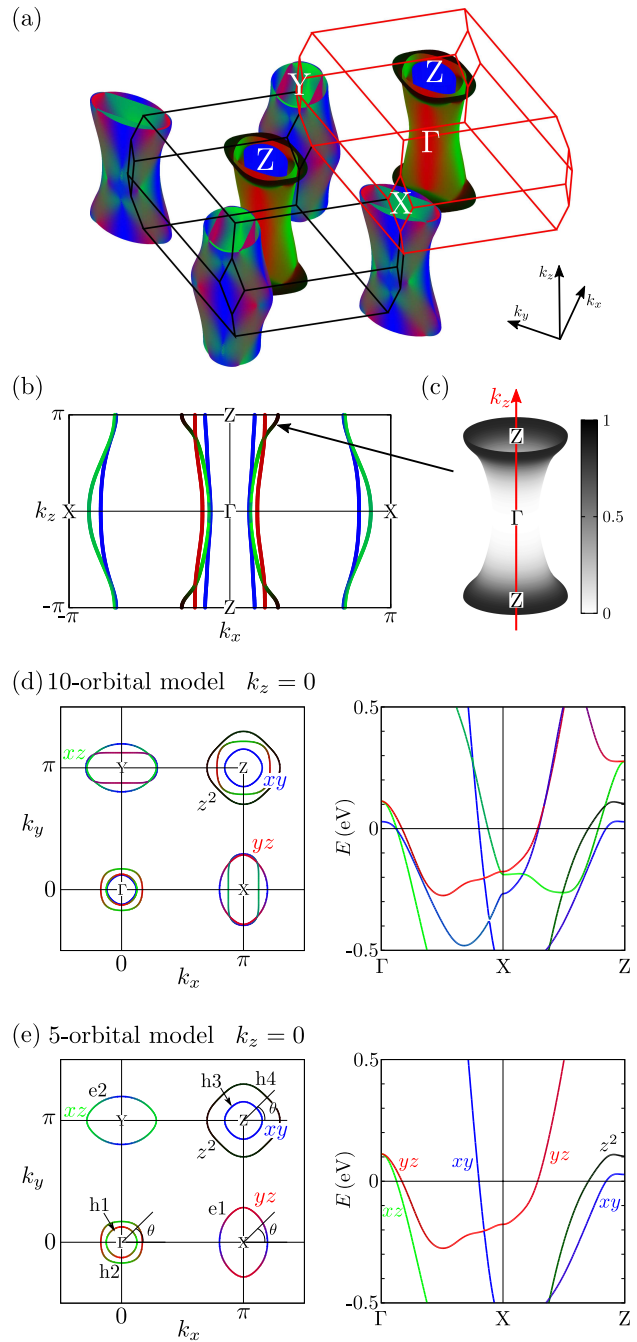}
\caption{(a) The three-dimensional FS of the optimally doped BaFe$_{2}$(As,P)$_{2}$. The solid lines show the Brillouin zone. Black, green, red and blue colors show the weight of the $z^{2}, {xz}, {yz}$ and ${xy}$ orbitals, respectively. (b) FS in the $k_{y}=0$ plane. (c) Schematic picture of the hole cylinder with the $k_{z}$ dependence of the weight of the $z^{2}$ orbital. (d) The FS and band structure of the ten-orbital model in the $k_{z}=0$ plane. Both ${\bm k}=(0,0,\pi)$ and ${\bm k}=(\pi,\pi,0)$ are the same Z-point. (e) The FS and band dispersions of the two-dimensional five-orbital model derived by unfolding the ten-orbital model in (d). The FS is essentially equal to that of LaFeAsO model except for the additional h4 composed of $z^{2}$ orbital.
}
\label{fig:model}
\end{figure}
In this paper, we introduce the two-dimensional five-orbital model for BaFe$_{2}$(As,P)$_{2}$ with the $3d$ orbitals $z^{2}, xz, yz, xy, x^{2}-y^{2}$ on Fe-ion (orbital 1-5). 
We derive the present model from the three-dimensional ten-orbital model for the optimally doped BaFe$_{2}$(As,P)$_{2}$ (30\% P-doped) introduced in Ref. \cite{Saito-loopnode}. 
Its three-dimensional FSs and Brillouin zone are shown in Fig.\,\ref{fig:model}(a). Black, green, red and blue colors show the weight of the $z^{2}, xz, yz$ and $xy$ orbitals. 
As shown in Fig.\,\ref{fig:model}(a), 
both ${\bm k}=(0,0,\pi)$ and ${\bm k}=(\pi,\pi,0)$ correspond to the same Z point
because BaFe$_{2}$(As,P)$_{2}$ has body-centered tetragonal structure.
Therefore, we can analyze the gap structure around Z point based on the two-dimensional model in the $k_{z}=0$ plane. 
The outer hole cylinder around Z point is composed of the $z^{2}$ orbital as shown in Figs.\,\ref{fig:model}(b) and (c).
Figure\,\ref{fig:model}(c) illustrates a schematic picture of the hole cylinder, on which the weight of the $z^{2}$ orbital is shown. 
The weight of the $z^{2}$ orbital is approximately 0.9 around Z point, and almost 0 around $\Gamma$ point. 
The gap structure on this hole cylinder around Z point is the main topic of this paper. 

The FS and band structure of the ten-orbital model in the $k_{z}=0$ plane are shown in Fig.\,\ref{fig:model}(d). 
Two electron FSs around X(Y) point are composed of the $xz, yz$ and $xy$ orbitals.
Three hole FSs around $\Gamma$ point and the inner and middle hole FSs around Z point are composed of the $xz, yz$ and $xy$ orbitals. 
The outer hole FS around Z point is composed of the $z^{2}$ orbital. By unfolding the ten-orbital model according to Refs. \cite{Kuroki,Miyake}, we derive the two-dimensional five-orbital model with the FSs and band structure shown in Fig.\,\ref{fig:model}(e). 
The hole FSs (h1,2) around $\Gamma$ point are composed of the $xz$ and $yz$ orbitals, and the hole FSs (h3,4) around Z point are composed of the 
$xy$ and $z^2$ orbitals. The electron FSs (e1,2) around X and Y points are composed of the $xz, yz$ and $xy$ orbitals.
The FS structure of the obtained BaFe$_{2}$(As,P)$_{2}$ model is essentially equivalent to that of LaFeAsO model in Ref. \cite{Onari-PRL2014} with the additional $z^{2}$ orbital hole FS (h4). 
We study the mechanism of superconductivity in the optimally doped BaFe$_{2}$(As,P)$_{2}$ based on the obtained two-dimensional five-orbital Hubbard model:
\begin{eqnarray}
H=H_{0}+H_{U}
\end{eqnarray}
The kinetic term $H_{0}$ is expressed as 
\begin{eqnarray}
H_{0}&=&\sum_{{\bm k},\sigma,l,m}H^{0}_{l,m}({\bm k})c^{\dagger}_{{\bm k},l\sigma}c_{{\bm k},m\sigma},
\end{eqnarray}
where ${\bm k}=(k_{x},k_{y}),\,\sigma=\uparrow$ or $\downarrow$ and $l,m=1-5$. $H^{0}_{l,m}({\bm k})$ is the two-dimensional five-orbital tight-binding model for BaFe$_{2}$(As,P)$_{2}$ shown in Fig.\,\ref{fig:model}(e),
which is given by putting $k_{z}=0$ in the original three-dimensional model.
$H_{U}$ is the multiorbital Coulomb interaction for the $d$-orbitals given as
\begin{eqnarray}
H_{U}=-\frac{1}{2}\sum_{i,ll',mm'}\sum_{\sigma\rho}U^{0}_{l\sigma,l'\sigma;m\rho,m'\rho}c^{\dagger}_{i,l\sigma}c_{i,l'\sigma}c^{\dagger}_{i,m'\rho}c_{i,m\rho}.
\end{eqnarray}
Here,
\begin{eqnarray}
U^{0}_{l\sigma,l'\sigma';m\rho,m'\rho'}&=&\frac{1}{2}U^{0c}_{l,l';m,m'}\delta_{\sigma,\sigma'}\delta_{\rho',\rho} \nonumber \\
&+&\frac{1}{2}U^{0s}_{l,l';m,m'}{\bm \sigma}_{\sigma,\sigma'}\cdot{\bm \sigma}_{\rho',\rho},
\end{eqnarray}
where ${\bm \sigma}=(\sigma_{x},\sigma_{y},\sigma_{z})$ is the Pauli matrix vector. 
${\hat U}^{0s}$ and ${\hat U}^{0c}$ are the bare Coulomb 
interaction matrices for spin and charge channels,
which are composed of intra-orbital Coulomb interaction $U$,
inter-orbital one $U'$, Hund's interaction $J$ 
and pair transfer $J'$ \cite{Yamakawa-FeSe}.
Here, we assume the relation $U=U'+2J$ and $J=J'$.
\begin{figure}[t]
\includegraphics[width=0.95\linewidth]{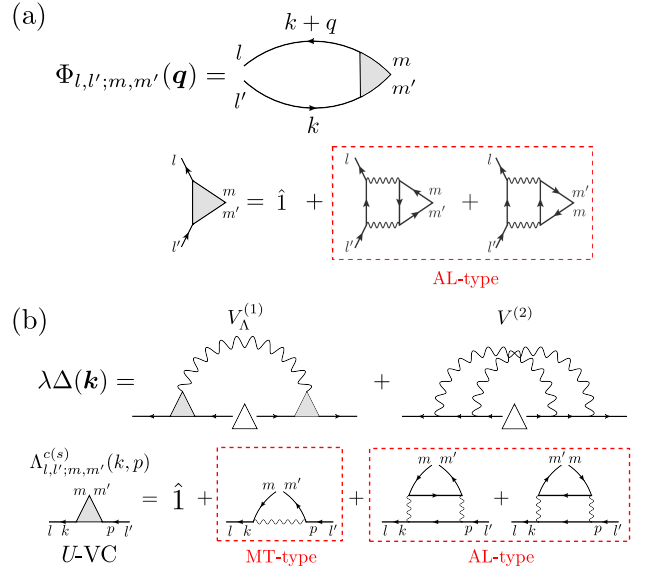}
\caption{(a) The irreducible susceptibility including the AL-VC. (b) The superconducting gap equation in the present theory. The {\it U}-VC composed of the MT- and AL-type VCs enhances (suppresses) the attractive (repulsive) term. $V^{(2)}$ term induces the attractive interaction in the present multiorbital model.
}
\label{fig:diagram}
\end{figure}

\section{orbital and spin susceptibilities}
\label{sec:susceptibility}
Based on the five-orbital Hubbard model for BaFe$_{2}$(As,P)$_{2}$, we calculate the orbital and spin susceptibilities 
based on the self-consistent VC (SC-VC) theory\cite{Onari-SCVC}.
In the SC-VC theory, we consider the AL-VC, which describes orbital-spin interference. 
Note that any VCs are ignored in the RPA. 
The irreducible susceptibility including the AL-VC is given in Ref. \cite{Yamakawa-PRB2017}.
\begin{eqnarray}
{\hat \Phi}^{c(s)}(q)=-T\sum_{k}{\hat G}(k+q){\hat G}(k)({\hat 1}+{\hat \Lambda}^{{\rm AL},c(s)}(k+q,k)), \label{eq:Phi}
\end{eqnarray}
where $k=({\bm k},\epsilon_{n})$ and $q=({\bm q},\omega_{l})$; $\epsilon_{n}=(2n+1)\pi T$ ($\omega_{l}=2l\pi T$) is the fermion (boson) Matsubara frequency.
${\hat G}(k)=[(i\epsilon_{n}+\mu){\hat 1}-{\hat H}_{0}]^{-1}$ is the Green function and $\mu$ is the chemical potential. 
The diagrammatic expression of ${\hat \Phi}^{c(s)}(q)$ is shown in Fig.\,\ref{fig:diagram}(a). 
The detailed expression of ${\hat \Lambda}^{{\rm AL},c(s)}$ is given in Ref. \cite{Yamakawa-PRB2017}. 
The charge (spin) susceptibility is given as
\begin{eqnarray}
{\hat \chi}^{c(s)}(q)={\hat \Phi}^{c(s)}(q)\left\{ {\hat 1}-{\hat U}^{0c(s)}{\hat \Phi}^{c(s)}(q)  \right\}^{-1}.
\end{eqnarray}
The charge (spin) susceptibility diverges when the charge (spin) Stoner factor $\alpha_{c(s)}$, which is given by the maximum eigenvalue of  ${\hat U}^{0c(s)}{\hat \Phi}^{c(s)}({\bm q},0)$, reaches unity. 
In the RPA analysis (${\hat \Lambda}^{{\rm AL},c(s)}=0$), the relation $\alpha_{s}>\alpha_{c}$ always holds for $J>0$ and $\chi^{c}(q)$ remains small even when $\chi^{s}(q)$ develops divergently. 
In contrast, the spin-fluctuation-driven orbital order or fluctuations are realized when we consider the VC since the AL-VC increases in 
proportion to $\sum_{p}\chi^{s}(q+p)\chi^{s}(p)$ near the magnetic QCP
\cite{Onari-SCVC}. 
We neglect the spin-channel AL-VC and Maki-Tompson VC (MT-VC) 
for susceptibilities since they are negligible in various models.
Figures \ref{fig:diagram}(b) shows the 
linearized gap equation with the $U$-VC.

\begin{figure}[t]
\includegraphics[width=0.8\linewidth]{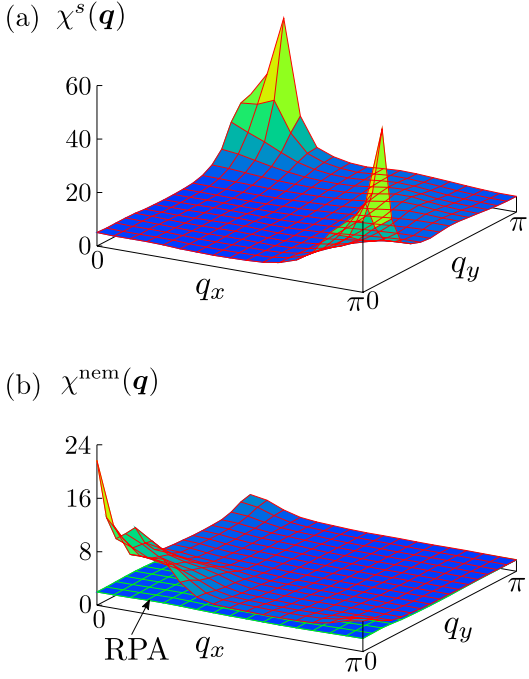}
\caption{(a)The spin susceptibility $\chi^{s}({\bm q})=\sum_{l,m}\chi^{s}_{l,l;m,m}({\bm q})$. (b) The nematic orbital susceptibility $\chi^{{\rm nem}}({\bm q})=\chi^{c}_{2,2;2,2}({\bm q})+\chi^{c}_{3,3;3,3}({\bm q})-\chi^{c}_{2,2;3,3}({\bm q})-\chi^{c}_{3,3;2,2}({\bm q})$ obtained by the SC-VC method. 
We also show $\chi^{{\rm nem}}({\bm q})$ obtained by RPA.
}
\label{fig:sc-vc}
\end{figure}
Hereafter, we carry out the calculation with $32\times  32\,{\bm k}$\,\,-mesh and 256 Matsubara frequencies. 
We fix the temperature at  $T=20$ meV and the ratio $J/U=0.1$. 
Figures\,\ref{fig:sc-vc}(a) and (b) are the spin and nematic orbital susceptibilities for $U=1.4$\,\,eV. 
The obtained spin and charge Stoner factors are $(\alpha_{s},\alpha_{c})=(0.97,0.88)$. 
The spin susceptibility $\chi^{s}({\bm q})=\sum_{l,m}\chi^{s}_{l,l;m,m}({\bm q})$ shows the maximum peak at ${\bm q}\simeq (\pi,0)$ due to the FS nesting, consistently with the magnetic order in under-doped compounds \cite{Ishikado,Ning,Kitagawa}. 
The nematic susceptibility $\chi^{{\rm nem}}({\bm q})=\chi^{c}_{2,2;2,2}({\bm q})+\chi^{c}_{3,3;3,3}({\bm q})-\chi^{c}_{2,2;3,3}({\bm q})-\chi^{c}_{3,3;2,2}({\bm q})$ shows the maximum peak at ${\bm q}=(0,0)$. 
The development of the nematic fluctuations is experimentally observed near the orthorhombic phase \cite{Fernandes-2010,Yoshizawa,Goto,Bohmer}. 
We also show the nematic orbital susceptibility given by RPA in Fig. \ref{fig:sc-vc} (b). 
Since it remains small, 
the structural phase transition cannot be explained by the RPA.

\begin{figure}[t]
\includegraphics[width=1.0\linewidth]{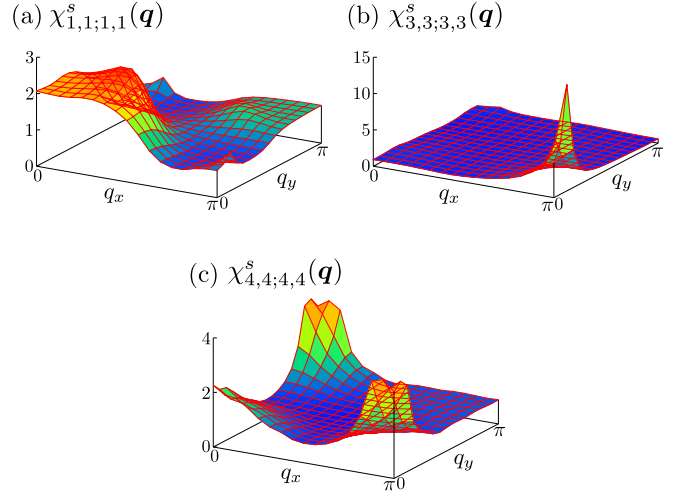}
\caption{The spin susceptibilities $\chi^{s}_{l,l;l,l}({\bm q})$ for (a) $l=1$, (b) $l=3$ and (c) $l=4$. $\chi^{s}_{1,1;1,1}({\bm q})$ moderately develops due to intra-FS nesting. $\chi^{s}_{3,3;3,3}({\bm q})$ and $\chi^{s}_{4,4;4,4}({\bm q})$ show large peaks at the inter-FS nesting vectors.
}
\label{fig:xs-orb}
\end{figure}

Figure\,\ref{fig:xs-orb} shows the intra-orbital spin susceptibilities, $\chi^{s}_{l,l;l,l}({\bm q})$, for (a) $l=1$, (b) $l=3$ and (c) $l=4$. $\chi^{s}_{1,1;1,1}({\bm q})$ shows the broad peak around ${\bm q}\sim {\bm0}$ due to the intra-FS nesting in h4.
There is no inter-FS nesting because of the absence of $z^{2}$ orbital 
weight in other FSs.
$\chi^{s}_{2,2;2,2}({\bm q})$ and $\chi^{s}_{3,3;3,3}({\bm q})$ are strongly enlarged due to nesting between h1,2 and e1,2.
Thus, spin fluctuations develop most strongly on $xz$ and $yz$ orbitals.
(Note that $\chi^{s(c)}_{2,2;2,2}(q_{x},q_{y})=\chi^{s(c)}_{3,3;3,3}(q_{y},q_{x})$.)
 $\chi^{s}_{4,4;4,4}({\bm q})$ is also enlarged due to the nesting between h3 and e1,2. 

\begin{figure}[htbp]
\includegraphics[width=1.0\linewidth]{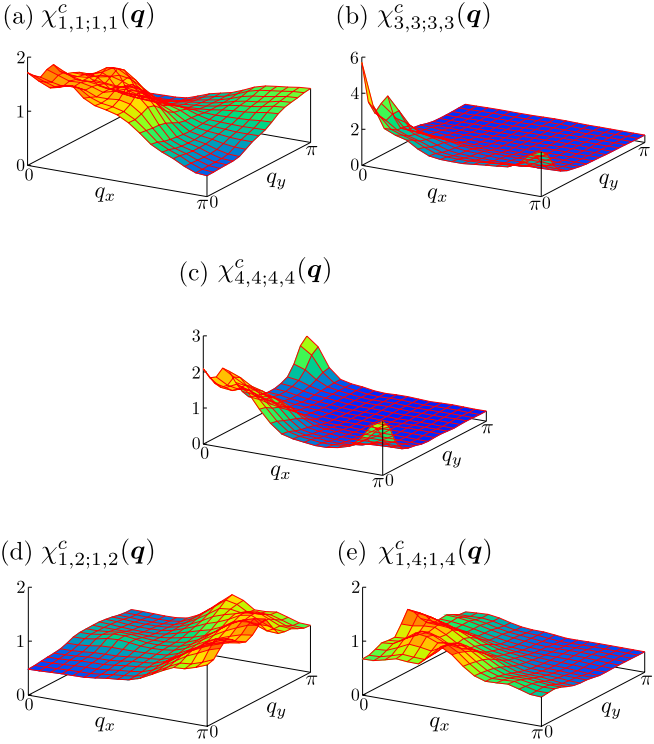}
\caption{The intra-orbital charge susceptibilities $\chi^{c}_{l,l;l,l}({\bm q})$ for (a) $l=1$, (b) $l=3$ and (c) $l=4$. $\chi^{c}_{1,1;1,1}({\bm q})$ shows the broad peak around ${\bm q}=(0,0)$. $\chi^{c}_{3,3;3,3}({\bm q})$ and $\chi^{c}_{4,4;4,4}({\bm q})$ exhibit both the ferro and anti-ferro peaks. The inter-orbital charge susceptibilities (d) $\chi^{c}_{1,2;1,2}({\bm q})$ and (c) $\chi^{c}_{1,4;1,4}({\bm q})$. These charge fluctuations are enlarged by the AL-VC.}
\label{fig:xc-orb}
\end{figure}

Figure\,\ref{fig:xc-orb} shows the intra-orbital charge susceptibilities, $\chi^{c}_{l,l;l,l}({\bm q})$, for (a) $l=1$, (b) $l=3$ and (c) $l=4$. $\chi^{c}_{1,1;1,1}({\bm q})$ shows the broad peak around ${\bm q}=(0,0)$, and $\chi^{c}_{3,3;3,3}({\bm q})$, $\chi^{c}_{4,4;4,4}({\bm q})$ show large ferro and antiferro fluctuations. 
The inter-orbital charge susceptibilities, $\chi^{c}_{1,2;1,2}({\bm q})$ and $\chi^{c}_{1,4;1,4}({\bm q})$, are also shown in Figs.\,\ref{fig:xc-orb}(d) and (e). These large charge fluctuations are caused by the AL-VC for the charge channel. 

\begin{figure*}[htbp]
\includegraphics[width=0.85\linewidth]{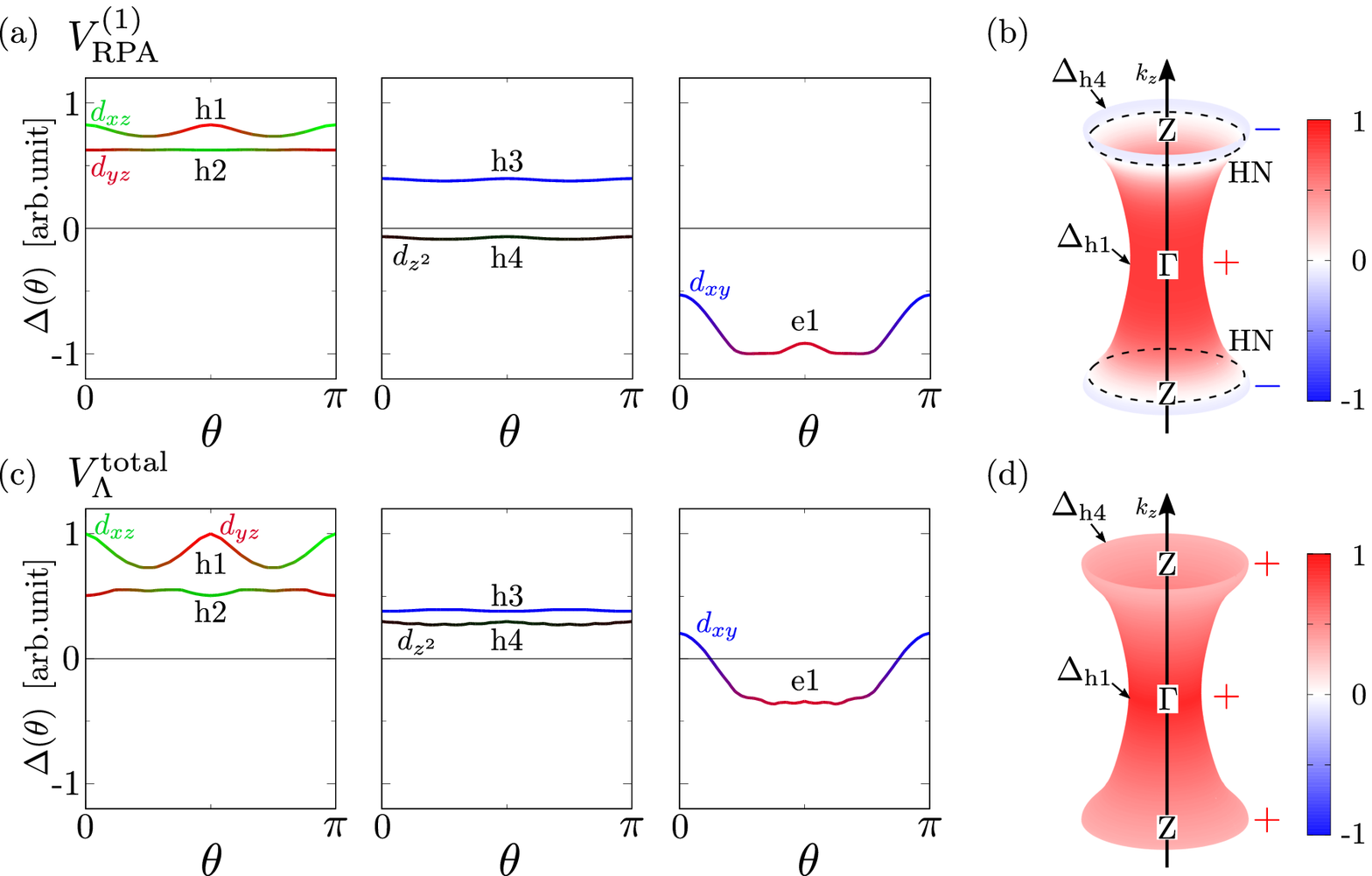}
\caption{(a) The fully-gapped $s_{+-}$ wave state given by $V^{(1)}_{{\rm RPA}}$. There is sign reversal between h4 and h1, that corresponds to the schematic horizontal node gap structure in (b). The broken lines represent the expected horizontal node. (c) The nodal $s$ wave state given by $V^{{\rm total}}_{\Lambda}$ (nearly $s_{+-}$). There is no sign reversal between h4 and h1, which corresponds to the schematic gap structure without horizontal node in (d). 
}
\label{fig:gap}
\end{figure*}
\section{gap equation beyond the Migdal-Eliashberg formalism}
\label{sec:gap}
\begin{figure}[htbp]
\includegraphics[width=1.0\linewidth]{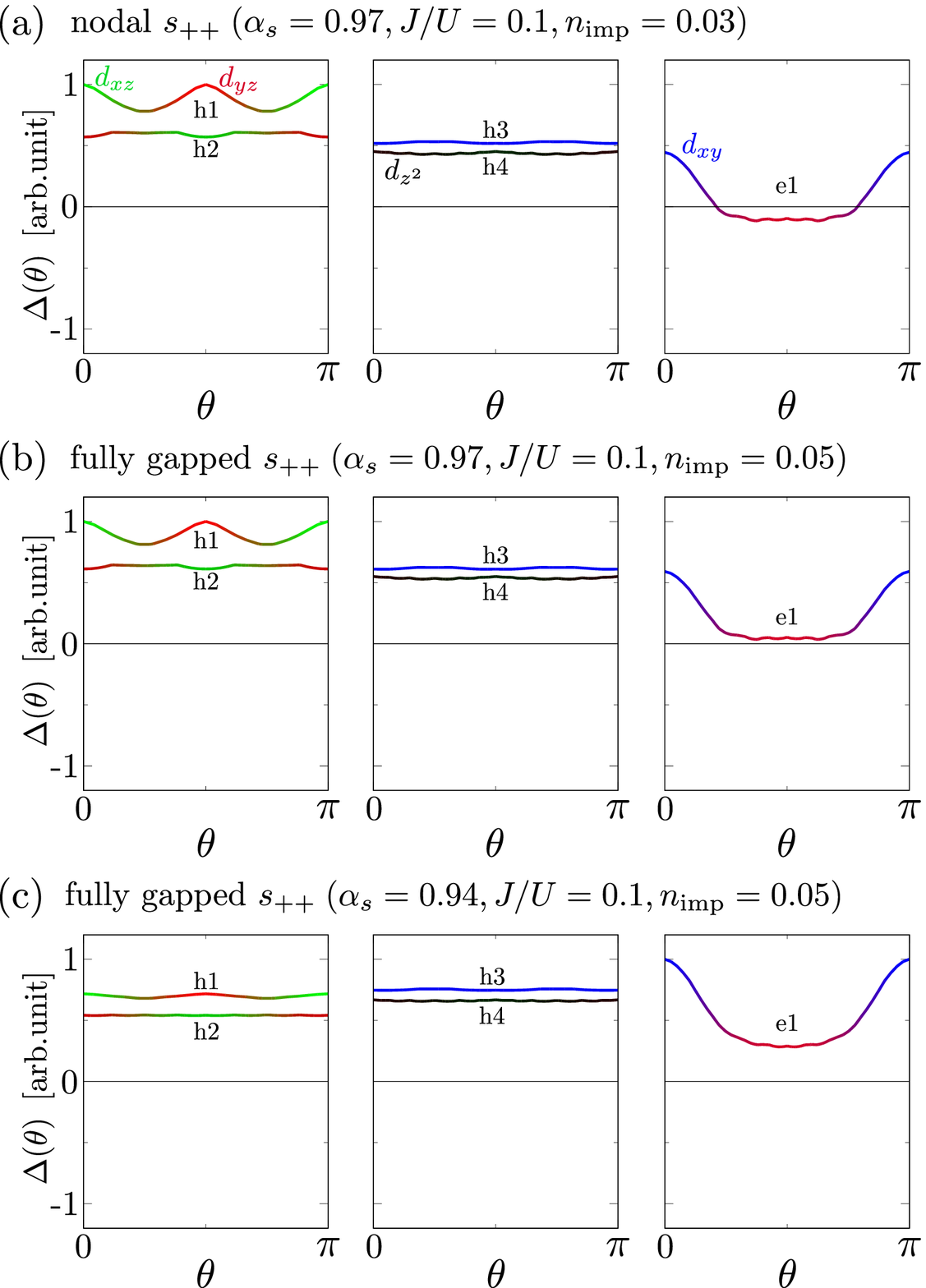}
\caption{The $\theta$ dependence of the gap structure. (a) The nodal $s$ wave state at $\alpha_{s}=0.97, J/U=0.1$ and $n_{{\rm imp}}=0.03$ (nearly $s_{++}$). (b) The fully-gapped $s_{++}$ wave state at $\alpha_{s}=0.97, J/U=0.1$ and $n_{{\rm imp}}=0.05$. (c) The perfectly fully-gapped $s_{++}$ wave state at $\alpha_{s}=0.94, J/U=0.1$ and $n_{{\rm imp}}=0.05$.
}
\label{fig:gapimp}
\end{figure}
\begin{figure*}[htbp]
\includegraphics[width=0.95\linewidth]{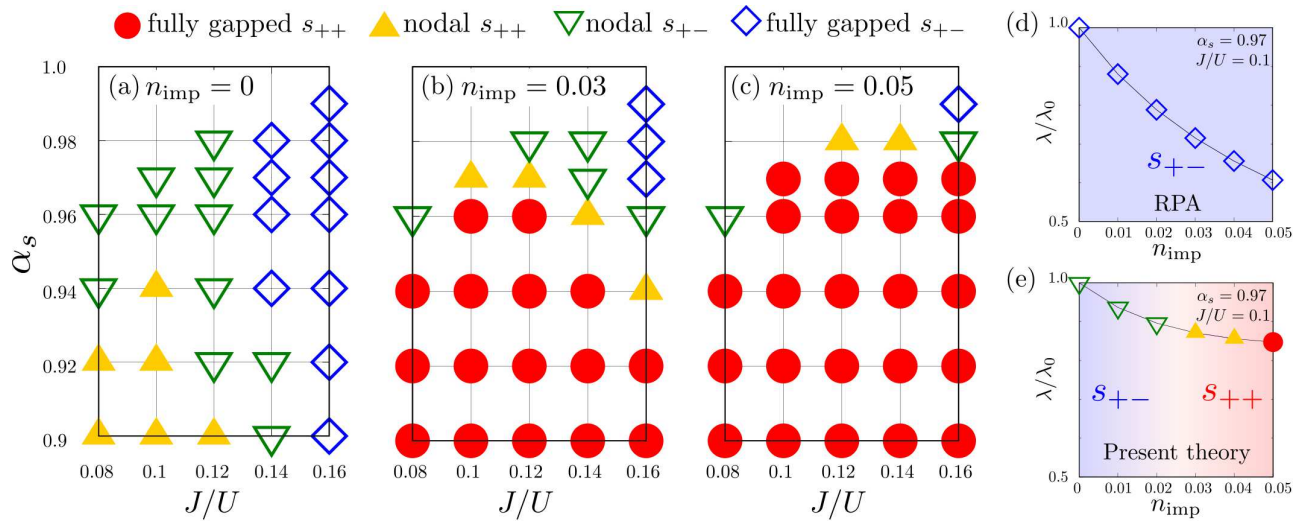}
\caption{(a) The $\alpha_{s}\mathchar`-J/U$ phase diagrams of the gap structure for (a) without impurity, (b) in the presence of the 3\% impurity, and (c) in the presence of the 5\% impurity, respectively. The red circles, yellow upper triangles, green lower triangles and blue squares show the fully-gapped $s_{++}$ wave state, nodal $s_{++}$ wave state ($\left<\Delta_{\rm e1}(k)\right>_{\rm FS}>0$), nodal $s_{+-}$ wave state ($\left<\Delta_{\rm e1}(k)\right>_{\rm FS}<0$) and fully-gapped $s_{+-}$ wave state, respectively. The impurity-dependence of $\lambda/\lambda_{0}$ for (d) RPA and (e) present theory with {\it U}-VC.
}
\label{fig:phase}
\end{figure*}

In this section,
we solve the linearized gap equation 
depicted in Fig. \ref{fig:diagram}(b),
concentrating on the superconducting gap function 
on the FSs \cite{Saito-loopnode}.
The equation is expressed as

\begin{eqnarray}
&&Z_{\alpha}({\bm k},\epsilon_{n})\lambda\Delta_{\alpha}({\bm k},\epsilon_{n}) \nonumber \\
&&=-\frac{\pi T}{(2\pi)^{2}}\sum_{\beta,m}\oint_{{\rm FS}_{\beta}}\frac{d{\bm p}}{v^{\beta}({\bm p})}
\left\{V^{{\rm pair}}_{\alpha,\beta}({\bm k},\epsilon_{n},{\bm p},\epsilon_{m})\right. \nonumber \\
&&\left.\hspace{3.9cm}+V^{\rm imp}_{\alpha,\beta}({\bm k},{\bm p};\epsilon_{n})\delta_{n,m}\right \}\nonumber \\
&&\hspace{3.9cm}\times\frac{\Delta_{\beta}({\bm p},\epsilon_{m})}{\left|\epsilon_{m}\right|},
\end{eqnarray}
where $\Delta_{\alpha}(k)$ is the superconducting gap function and $\alpha,\beta$ are indices of FS.
The eigenvalue $\lambda$ is approximately proportional to $T_{\rm c}$ and $\lambda=1$ is satisfied at $T=T_{\rm c}$. $V_{\alpha,\beta}^{{\rm pair}}(k,p)$ is the pairing interaction in the band-diagonal basis given as
\begin{eqnarray}
&&V_{\alpha,\beta}^{{\rm pair}}(k,p)=\sum_{ll'mm'}V^{{\rm pair}}_{l,l';m,m'}(k,p) \nonumber \\
&&\hspace{1cm}\times u^{\ast}_{l\alpha}({\bm k})u_{l'\beta}({\bm p})u_{m\beta}(-{\bm p})u^{\ast}_{m'\alpha}(-{\bm k}),
\end{eqnarray}
where $u_{l\alpha}({\bm k})=\left<l|{\bm k};\alpha\right>$ is the unitary matrix that connects between band basis and orbital basis.
(The expression of $V^{{\rm pair}}(k,p)$ will be presented
at the end of this section; see Eq.(\ref{eq:vtotal}).
We consider the impurity effect based on the $T$ matrix approximation.
$V^{{\rm imp}}({\bm k},{\bm p};\epsilon_{n})$, which is induced by impurities, is given as
\begin{eqnarray}
&&V^{{\rm imp}}_{\alpha,\beta}({\bm k},{\bm p};\epsilon_{n})=-\frac{n_{\rm imp}}{T}\sum_{ll'mm'}T_{ll'}(\epsilon_{n})T_{mm'}(-\epsilon_{n}) \nonumber \\
&&\hspace{1cm}\times u^{\ast}_{l\alpha}({\bm k})u_{l'\beta}({\bm p})u_{m\beta}(-{\bm p})u^{\ast}_{m'\alpha}(-{\bm k}),
\end{eqnarray}
where $n_{\rm imp}$ is impurity concentration.
We consider the diagonal impurity potential $I_{\rm imp}$ in the $d$ orbital basis. 
The $T$ matrix for an impurity is given as
\begin{eqnarray}
{\hat T}(\epsilon_{n})=\left[{\hat 1}-{\hat I}{\hat G}_{\rm loc}(\epsilon_{n})\right]^{-1}{\hat I}.
\end{eqnarray}
Here, $I_{ll'}=I_{\rm imp}\delta_{l,l'}$, and $[G_{\rm loc}(\epsilon_{n})]_{ll'}=\displaystyle{\sum\limits_{\bm k}}G_{ll'}({\bm k},\epsilon_{n})$ is the local Green function. 
The normal self-energy induced by impurities is given as
\begin{eqnarray}
\delta\textstyle{\sum_{\alpha}^{n}}({\bm k},\epsilon_{n})=n_{\rm imp}\sum_{ll'}u^{\ast}_{l\alpha}({\bm k})T_{ll'}(\epsilon_{n})u_{l'\alpha}({\bm k}).
\end{eqnarray}
Then, $Z_{\alpha}({\bm k},\epsilon_{n})$ is given as
\begin{eqnarray}
Z_{\alpha}({\bm k},\epsilon_{n})=1+\frac{\gamma^{\alpha}({\bm k},\epsilon_{n})}{\left|\epsilon_{n}\right|}, 
\end{eqnarray}
where $\gamma^{\alpha}({\bm k},\epsilon_{n})=-{\rm Im}\,\delta\textstyle{\sum_{\alpha}^{n}}({\bm k},\epsilon_{n}){\rm sgn}(\epsilon_{n})$ is the impurity-induced quasiparticle damping rate.

Beyond the ME formalism, we take the {\it U}-VC for the coupling constant into account. 
The MT-VC and AL-VC for {\it U}-VC are 
depicted in Fig. \ref{fig:diagram}(b),
and their analytic expressions are given in Ref. \cite{Yamakawa-PRB2017}.
The total $U$-VC for the charge (spin) channel is 
\begin{eqnarray}
{\hat \Lambda}^{c(s)}(k,k')={\hat 1}+{\hat \Lambda}^{{\rm MT},c(s)}(k,k')+{\hat \Lambda}^{{\rm AL},c(s)}(k,k').
\end{eqnarray}
The effect of the spin-channel AL-VC on ${\hat \chi}^{s}({\bm q})$ is small
since $\left|\Lambda^{{\rm AL},s(c)}(k+q,k) \right|\ll1$ except for 
small Matsubara frequencies \cite{Yamakawa-PRB2017}. 
In contrast, the AL-VC in the gap equation is important
since Cooper pairs are formed by low-energy quasiparticles.
By taking the {\it U}-VC into account, the single fluctuation exchange term 
in the pairing interaction is given as
\begin{eqnarray}
{\hat V}^{(1)}_{\Lambda}(k,p)&=&{\hat V}^{(1)s}_{\Lambda}(k,p)+{\hat V}^{(1)c}_{\Lambda}(k,p)+{\hat V}^{0},
\end{eqnarray}
where
\begin{eqnarray}
{\hat V}^{(1)s}_{\Lambda}(k,p)&=&\frac{3}{2}{\hat V}^{\Lambda,s}(k,p) \label{eq:vs} \\
{\hat V}^{(1)c}_{\Lambda}(k,p)&=&-\frac{1}{2}{\hat V}^{\Lambda,c}(k,p) \label{eq:vc} \\
{\hat V}^{0}&=&-{\hat U}^{0s} \label{eq:double}.
\end{eqnarray}
Equations\,(\ref{eq:vs}) and (\ref{eq:vc}) represent the 
spin- and charge-fluctuation-mediated interaction terms with $U$-VCs,
and Eq. (\ref{eq:double}) is necessary to eliminate the double counting. 
Here, ${\hat V}^{\Lambda,c(s)}$ is given as
\begin{eqnarray}
{\hat V}^{\Lambda c(s)}(k,p)={\hat \Lambda}^{c(s)}(k,p){\hat V}^{c(s)}(k-p){\hat {\bar \Lambda}}^{c(s)}(-k,-p),
\end{eqnarray}
where
\begin{eqnarray}
{\hat V}^{c(s)}(k-p)&=&{\hat U}^{0c(s)}+{\hat U}^{0c(s)}{\hat \chi}^{c(s)}(k-p){\hat U}^{0c(s)} \label{eq:vcs}\\
{\hat {\bar \Lambda}}_{l,l';m,m'}^{c(s)}(k,p)&=&{\hat \Lambda}^{c(s)}_{m,m';l,l'}(k,p).
\end{eqnarray}
The pairing interaction due  to the charge (spin) fluctuations is enhanced (suppressed) by the {\it U}-VC since $|\Lambda^{c(s)}(k,p)|^{2}$ is larger (smaller) than 1. 

To discuss the importance of the $U$-VC,
we also study the pairing interaction given by RPA
without $U$-VC ($\hat{\Lambda}^{c,s}=\hat{1}$) in later sections:
\begin{eqnarray}
{\hat V}^{(1)}_{{\rm RPA}}(k-p)={\hat V}^{(1)s}_{{\rm RPA}}(k-p)+{\hat V}^{(1)c}_{{\rm RPA}}(k-p)+{\hat V}^{0}.
\end{eqnarray}
In this ME formalism, ${\hat V}^{(1)s}_{{\rm RPA}}$ is important because only spin fluctuations develop in the RPA.

In addition, we calculate the double fluctuation exchange pairing interaction $V^{(2)}$ term, which corresponds to the AL process for the pairing interaction.
It is given as
\begin{eqnarray}
V^{(2)}_{l,l';m,m'}(k,p)&=&\frac{T}{4}\sum_{q}\sum_{a,b,c,d}G_{a,b}(p-q)G_{c,d}(-k-q) \nonumber \\
&&\times\{3V^{s}_{l,a,m,d}(k-p+q)V^{s}_{b,l',c,m'}(-q) \nonumber \\
&&+3V^{s}_{l,a,m,d}(k-p+q)V^{c}_{b,l',c,m'}(-q) \nonumber \\
&&+3V^{c}_{l,a,m,d}(k-p+q)V^{s}_{b,l',c,m'}(-q) \nonumber \\
&&-V^{c}_{l,a,m,d}(k-p+q)V^{c}_{b,l',c,m'}(-q)\} \label{eq:Vcross}
\end{eqnarray}
In the present model, $V^{(2)}$ induces attractive 
(repulsive) interaction for 
${\bm k}-{\bm p}\approx (\pi,0)$
(${\bm k}-{\bm p}\approx (0,0)$)
as we discussed in Ref. \cite{Yamakawa-PRB2017} in detail.
The total pairing interaction 
in the present beyond ME formalism is given as
\begin{eqnarray}
{\hat V}^{{\rm total}}_{\Lambda}(k,p)={\hat V}^{(1)}_{\Lambda}(k,p)+{\hat V}^{(2)}(k,p). \label{eq:vtotal}
\end{eqnarray}

\section{The $s_{++}$ wave state without horizontal node}
Hereafter, we discuss the obtained gap functions at $\alpha_{s}=0.97$ and $J/U=0.1$.
Figure\,\ref{fig:gap}(a) is the gap function derived from ${\hat V}^{(1)}_{{\rm RPA}}$, which is the pairing interaction in the ME formalism.
This is the fully-gapped $s_{+-}$ wave state with very small $|\Delta_{\rm h4}|$: 
The sign reversal between h4 and h1 corresponds to the presence of the horizontal node,
which is expressed by the schematic gap structure in Fig.\,\ref{fig:gap}(b). 
The broken lines represent the expected horizontal node. 
This result is consistent with the previous RPA results in Ref. \cite{Suzuki-Kuroki}. 
However, both the ARPES studies in Refs. \cite{Shimojima,Yoshida} and the small Volovik effect in the specific heat measurement in Refs. \cite{Kim,Wang} indicate the absence of horizontal node.

Figure\,\ref{fig:gap}(c) is the gap function derived from 
${\hat V}^{{\rm total}}$.
In this case, nodal $s$ wave state is obtained.
There is no sign reversal between h4 and h1, that corresponds to the absence of the horizontal node expressed by the schematic gap structure in Fig.\,\ref{fig:gap}(d). 
This result is consistent with the ARPES studies in Refs. \cite{Shimojima,Yoshida}. 
We call it the nodal $s_{+-}$ wave state, since the gap on e1 is mainly negative, that is, $\left<\Delta_{\rm e1}(k)\right>_{\rm FS}<0$. When we neglect the {\it U}-VC in the gap equation, 
we obtain the $s_{+-}$ wave state essentially similar to the gap structure in Fig.\,\ref{fig:gap}(a).
This fact means that the {\it U}-VC must be included in the gap equation
to obtain reliable results.

In Fig.\,\ref{fig:gap}(c), nodes appear only on the electron-FSs,
at which the orbital character gradually changes between $xz(yz)$ and $xy$.
This result means the emergence of the loop-nodes on the electron-FSs,
consistently with the theoretical prediction in Ref. \cite{Saito-loopnode} and the angle-resolved
thermal conductivity measurement \cite{Yamashita}.

\subsection{The $J/U$ and impurity dependences of the gap structure}
As shown in Fig.\,\ref{fig:gap}(c), we obtained the nodal $s$ wave state with large gap on h4 by applying the present beyond ME formalism. 
Here, we discuss impurity effect on the gap structure.
We consider the on-site intra-orbital impurity potential 
with $I_{\rm imp}=1$\,eV. 
Figure\,\ref{fig:gapimp}(a) shows the obtained gap function for $\alpha_{s}=0.97$ and $J/U=0.1$ in the presence of the 3\% impurities $(n_{\rm imp}=0.03)$.
The gap on e1 is mainly positive ($\left<\Delta_{\rm e1}(k)\right>_{\rm FS}>0$), so we call it the nodal $s_{++}$ wave state.
With increasing $n_{\rm imp}$, we obtain the fully-gapped $s_{++}$ wave state,
as shown in Fig.\,\ref{fig:gapimp}(b) for $n_{{\rm imp}}=0.05$. 
Thus, impurity-induced $s_{+-} \rightarrow s_{++}$ crossover is realized in the present study \cite{Kontani-cross,Hirschfeld-cross}.

The gap structure sensitively depends on the Stoner factor
and model parameters.
In Fig.\,\ref{fig:gapimp}(c), we show the gap structure obtained for 
$\alpha_{s}= 0.94$ and $n_{\rm imp}=0.05$.
The obtained fully-gapped $s_{++}$ wave function becomes more isotropic
compared with the gap in Fig.\,\ref{fig:gapimp}(b).

Next, we show the $\alpha_{s}\mathchar`-J/U$ phase diagrams of the gap structure in Fig.\,\ref{fig:phase} in the case of (a) $n_{{\rm imp}}=0$, (b) $n_{{\rm imp}}=0.03$ and (c) $n_{{\rm imp}}=0.05$. 
The red circles, yellow upper triangles, green lower triangles and blue squares represent the fully-gapped $s_{++}$ wave, nodal $s_{++}$ wave ($\left<\Delta_{\rm e1}(k)\right>_{\rm FS}>0$), nodal $s_{+-}$ wave ($\left<\Delta_{\rm e1}(k)\right>_{\rm FS}<0$), and fully-gapped $s_{+-}$ wave states, respectively. 
In Fig.\,\ref{fig:phase}(a), both $s_{++}$ wave and $s_{+-}$ wave states appear.
In contrast, in Fig.\,\ref{fig:phase}(b), the $s_{++}$ wave state is realized in the wide parameter range. 
The range of the $s_{++}$ wave state is expanded further by increasing the impurity concentration, as shown in Fig.\,\ref{fig:phase}(c). 
The phase diagrams (a)-(c) show that the crossover between the $s_{+-}$ wave state and $s_{++}$ wave state is caused by a small amount of impurities. 

To discuss the impurity effect on $T_{{\rm c}}$, we show the impurity dependence of the normalized eigenvalue, $\lambda/\lambda_{0}$, where
$\lambda_{0}$ is the eigenvalue at $n_{\rm imp}=0$.
Figures \ref{fig:phase} (d) and (e) show the obtained $\lambda/\lambda_{0}$ given by RPA and the present theory, respectively.
$\lambda_{0}=1.4$ in the RPA analysis, and $\lambda_{0}=1.2$ in the present theory.
In the RPA, $\lambda/\lambda_{0}$ decreases drastically, which indicates the drastic reduction in $T_{{\rm c}}$. 
In contrast, the decrease of $\lambda/\lambda_{0}$ given by the present theory is small, which indicates the small reduction in $T_{{\rm c}}$ during the $s_{+-}\rightarrow s_{++}$ crossover \cite{Kontani-cross,Hirschfeld-cross}. 
In fact, the robustness of $T_{\rm c}$ against impurities is confirmed in many Fe-based superconductors \cite{Sato-imp,Nakajima,Li,Paglione}.
We expect that the $s_{++}$ wave state is realized in real compounds due to small amount of impurities.

\subsection{The mechanism of the $s_{++}$ wave state with the absence of horizontal node on the FS composed of the $z^{2}$ orbital}
\begin{figure}[htbp]
\includegraphics[width=0.95\linewidth]{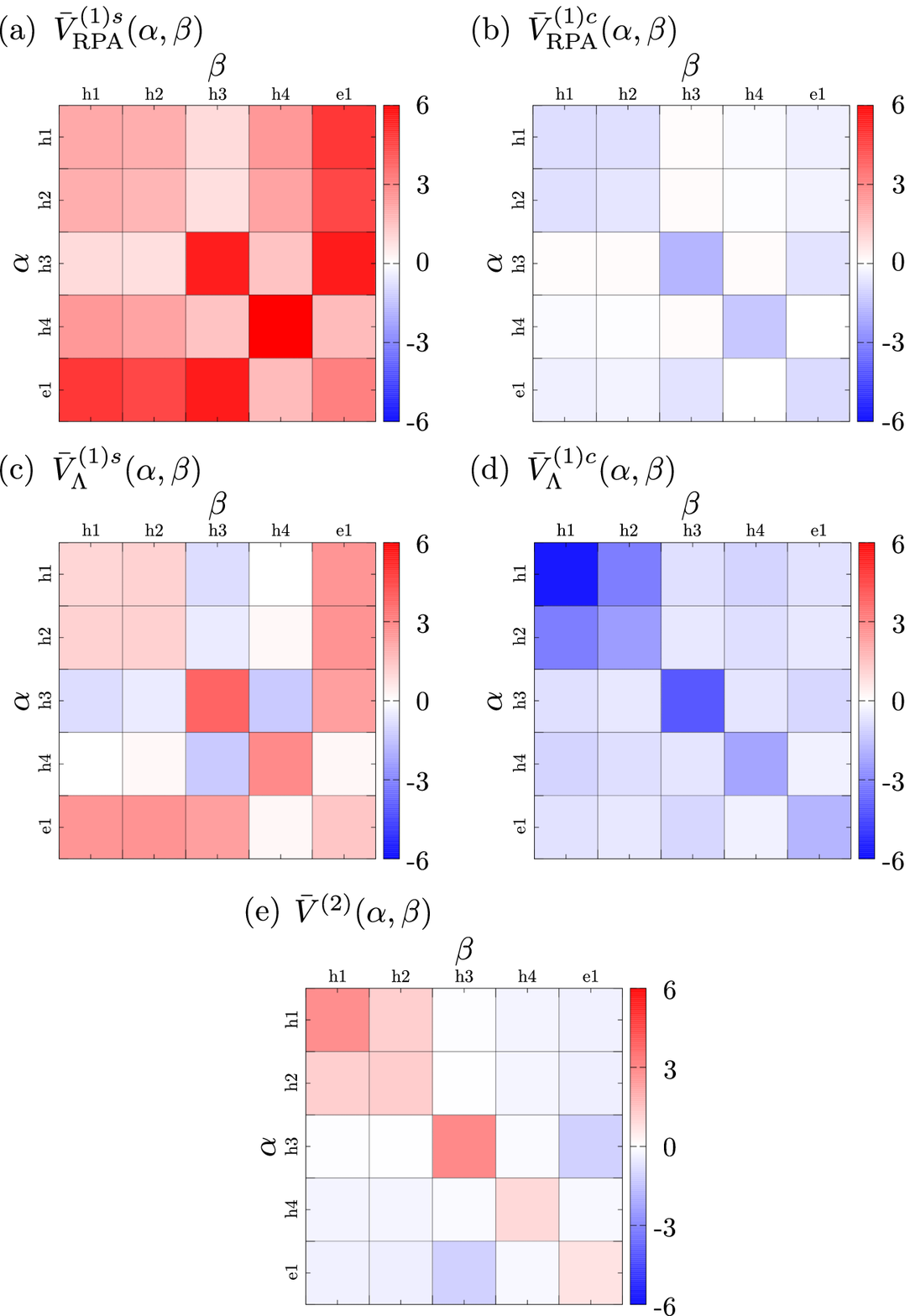}
\caption{The averaged pairing interaction between FS$_{\alpha}$ and FS$_{\beta}$ for (a) $V^{(1)s}_{{\rm RPA}}$, (b) $V^{(1)c}_{{\rm RPA}}$, (c) $V^{(1)s}_{\Lambda}$, (d) $V^{(1)c}_{\Lambda}$, and (e) $V^{(2)}$. Blue and red panels show that interaction between FSs is attractive and repulsive, respectively. Here, we drop the bare Coulomb repulsive term, which is reduced by retardation effect.
}
\label{fig:vpair}
\end{figure}
\begin{figure}[t]
\includegraphics[width=1.0\linewidth]{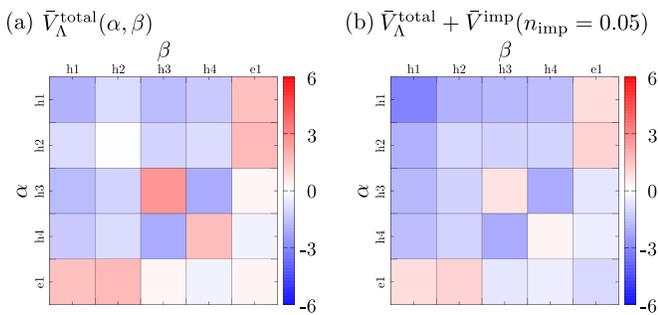}
\caption{The averaged $V^{{\rm total}}$ between FS$_{\alpha}$ and FS$_{\beta}$ for (a) without impurity and (b) in the presence of the 5\% impurities. 
}
\label{fig:vpair_total}
\end{figure}
\begin{figure}[htbp]
\includegraphics[width=0.7\linewidth]{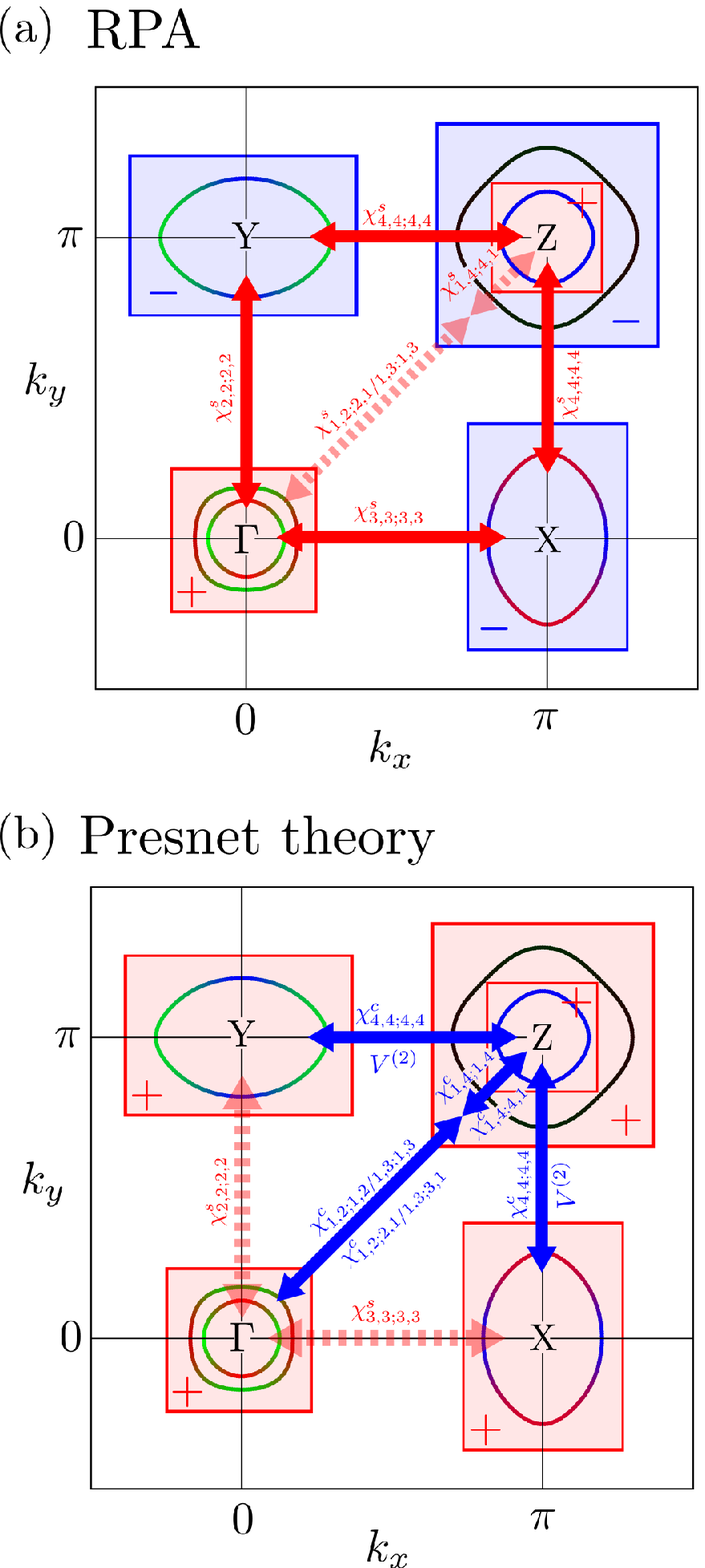}
\caption{The schematic pairing interaction between FSs: (a) In RPA analysis. (b) In the present theory. The blue and red arrows represent attractive and repulsive interactions, respectively. The broken arrows show weak interaction.
}
\label{fig:schematic}
\end{figure}

Here, we explain the reason why the $s_{++}$ wave state
is obtained in the present study.
To clarify the pairing interaction between FSs, we define the averaged pairing interaction between $\alpha$ and $\beta$ as
\begin{eqnarray}
{\bar V}(\alpha,\beta)=\frac{\oint d{\bm k}_{\alpha} d{\bm k}_{\beta} V^{\rm pair}_{\alpha,\beta}({\bm k}_{\alpha},{\bm k}_{\beta})}{\oint d{\bm k}_{\alpha} d{\bm k}_{\beta}},
\end{eqnarray}
where ${\bm k}_{\alpha}$ is Fermi wavenumber on FS denoted by $\alpha$. 
We show ${\bar V}(\alpha,\beta)$ in Fig.\,\ref{fig:vpair};(a) $V^{\rm pair}_{\alpha,\beta}=V^{(1)s}_{{\rm RPA}}$, (b) $V^{(1)c}_{{\rm RPA}}$, (c) $V^{(1)s}_{\Lambda}$, (d) $V^{(1)c}_{\Lambda}$, and (e) $V^{(2)}$. 
Blue and red panels mean that interaction between FSs is attractive and repulsive, respectively.
In the RPA analysis, repulsive interaction by $V^{(1)s}_{{\rm RPA}}$ is dominant. 
The repulsive interaction is strong between h1-3 and e1.
The repulsive interaction is given by intra-orbital spin fluctuations ($\chi^{s}_{2,2;2,2},\chi^{s}_{3,3;3,3}$ and $\chi^{s}_{4,4;4,4}$), and induces the $s_{+-}$ wave state. 
In contrast, the inter-FS repulsive interaction between h4 and the other FSs is weak, as shown in panels (h4,$\beta$) for $\beta \neq {{\rm h4}}$ in Fig.\,\ref{fig:vpair}(a).
Small negative gap in h4 is induced by weak repulsive interactions due to inter-orbital spin fluctuations ($\chi^{s}_{1,2;2,1},\chi^{s}_{1,3;3,1}$ and $\chi^{s}_{1,4;4,1}$). 

In highly contrast, ${\bar V}^{(1)c}_{\Lambda}$ shown in Fig.\,\ref{fig:vpair}(d) is strongly attractive because of the large charge-channel {\it U}-VC: $|\Lambda^{c}_{l,l;l,l}|^{2}\gg 1$. 
Especially, strong attractive interaction in the panels $({{\rm h1},{\rm h1}})$, $({{\rm h1},{\rm h2}})$ and $({{\rm h2},{\rm h2}})$ originates from the ferro-orbital fluctuations at ${\bm q}=(0,0)$ shown in Fig.\,\ref{fig:sc-vc}(a). 
Thus, nematic fluctuations are significant for the pairing mechanism. 
${\bar V}^{(1)s}_{\Lambda}$ with the {\it U}-VC for spin channel shown in Fig.\,\ref{fig:vpair}(c) gives weak repulsive interaction compared with ${\bar V}^{(1)s}_{{\rm RPA}}$, because spin fluctuation mediated pairing interactions are reduced by $|\Lambda^{s}_{l,l;l,l}|^{2}< 1$.
Interestingly, ${\bar V}^{(1)s}_{\Lambda}(\alpha,\beta)$ for $(\alpha,\beta)$=(h1,h3),(h2,h3) and (h3,h4) is weakly attractive,
if the $U$-VC is taken into account.
The reason is shortly explained in Appendix \ref{sec:appendix}. 
Therefore, {\it U}-VCs for both spin and charge channels play important role for realizing the $s_{++}$ wave state, whereas the $s_{+-}$ wave state is suppressed.
Moreover, $V^{(2)}$ gives important contribution to the $s_{++}$ wave state. 
In fact, both ${\bar V}^{(2)}({{\rm e1},{\rm h1\mathchar`-4}})$ and ${\bar V}^{(2)}({{\rm h4},{\rm h1,2}})$ give attractive interaction between different FSs.
In contrast, ${\bar V}^{(2)}$ gives intra-FS repulsive interaction.
Note that $|V^{(2)}|$ is large when FSs have large $xy$ orbital component, like ${\bar V}^{(2)}({{\rm e1},{\rm h3}})$. 

${\bar V}^{{\rm total}}_{\Lambda}$ without impurity is shown in Fig.\,\ref{fig:vpair_total}(a). 
This total pairing interaction gives the nodal $s$ wave state shown in Fig.\,\ref{fig:gap}(c).
The inter-FS and intra-FS interactions in Fig.\,\ref{fig:vpair_total}(a), which are attractive or repulsive depending on FSs, 
are averaged by introducing impurities.
For this reason the number of red panels in Fig.\,\ref{fig:vpair_total}(b) for $n_{\rm imp}=0.05$, which shows ${\bar V}^{{\rm total}}_{\Lambda}+{\bar V}^{\rm imp}$,
is smaller than that in Fig.\,\ref{fig:vpair_total}(a).
The pairing interaction at $n_{\rm imp}=0.05$ in Fig.\,\ref{fig:vpair_total}(b) 
gives the fully-gapped $s_{++}$ wave state in Fig.\,\ref{fig:gapimp}(c).
Therefore, by introducing small amount of impurities,
the attractive interaction becomes dominant 
and therefore $s_{++}$ wave state is realized. 

Next, we focus on the gap function in h4: We explain the reason why $\Delta_{{\rm h4}}$ is large and $\Delta_{{\rm h4}}\Delta_{{\rm h1}}>0$ is realized as shown in Fig.\,\ref{fig:gap}(c): This result means that the horizontal node is absent. 
In Figs.\,\ref{fig:vpair_total}(a) and (b), we find that ${\bar V}^{{\rm total}}$ is attractive between h4 and other hole FSs. 
The attractive interaction between h4 and h1-3 is given by inter-orbital fluctuations $\chi^{c}_{1,m;m,1}$ and $\chi^{c}_{1,m;1,m}(m=2-4)$, and they are strongly enlarged by the inter-orbital {\it U}-VC. 

We summarize the schematic pairing interactions between FSs given by RPA in Fig.\,\ref{fig:schematic}(a), and those given by the present theory with {\it U}-VCs in Fig.\,\ref{fig:schematic}(b). 
In the RPA, strong repulsive interactions driven by intra-orbital spin fluctuations are dominant, and the $s_{+-}$ wave state is realized. 
The horizontal node appears due to weak repulsive interactions given by inter-orbital spin fluctuations \cite{Suzuki-Kuroki}. 
In the present theory with {\it U}-VCs, the $s_{++}$ wave state is realized, because the attractive interactions by orbital fluctuations are strongly enhanced 
whereas repulsive interactions by spin fluctuations are reduced by the {\it U}-VCs. 
$V^{(2)}$ also induces attractive inter-pocket interaction. 
In addition, $\Delta_{{\rm h4}}$ is large and the relation $\Delta_{{\rm h1}} \simeq \Delta_{{\rm h2}} \simeq \Delta_{{\rm h3}} \simeq \Delta_{{\rm h4}}$ holds, due to the inter-orbital fluctuations as discussed in Ref. \cite{Saito-loopnode}. Therefore, horizontal node is absent in the present beyond ME theory.

\subsection{Anisotropy of the gap structure}
\begin{figure}[htbp]
\includegraphics[width=0.95\linewidth]{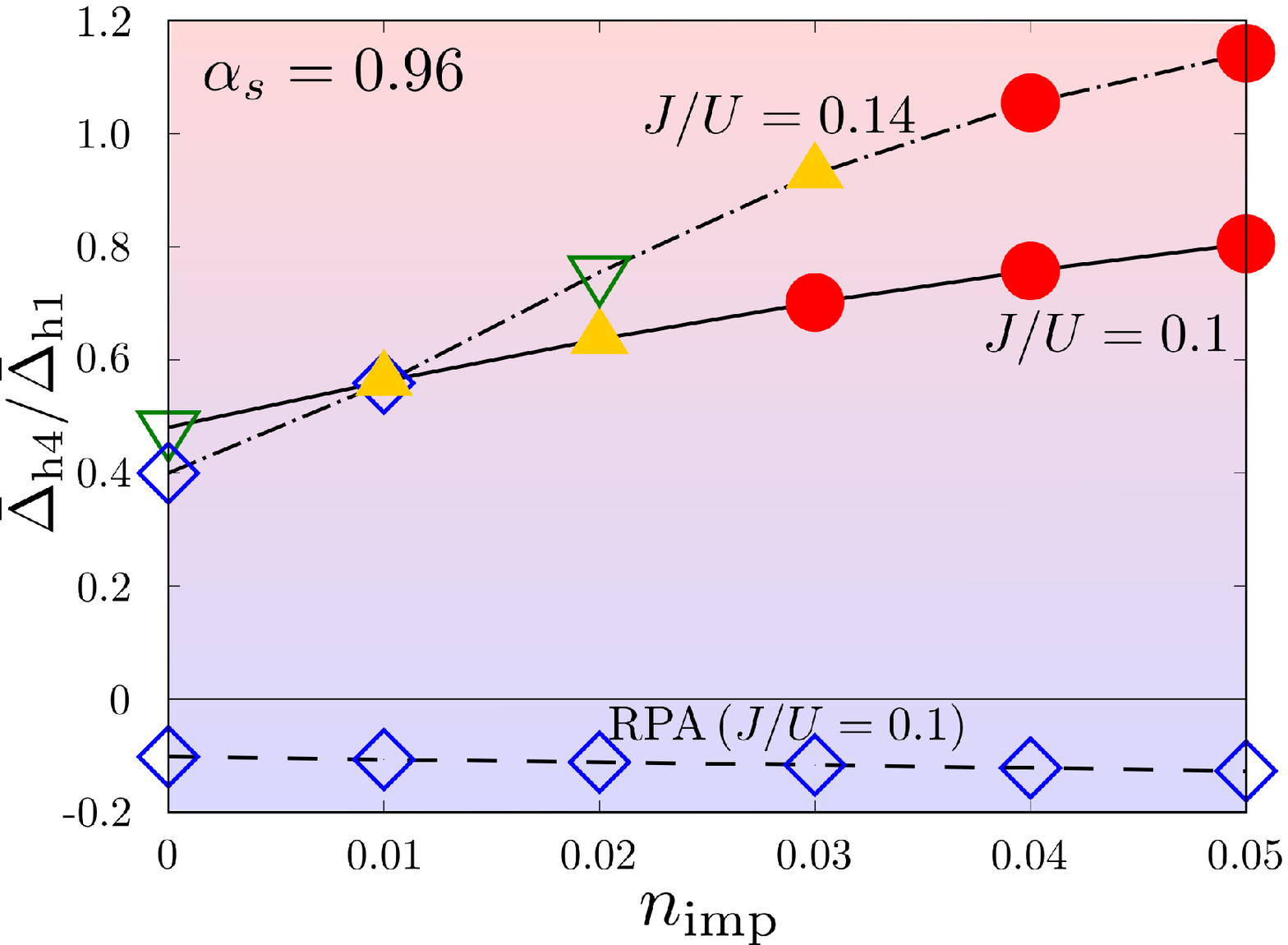}
\caption{The impurity-dependence of the averaged gap size ratio ${\bar \Delta}_{{\rm h4}}/{\bar \Delta}_{{\rm h1}}$ for $\alpha_{s}=0.96$.  ${\bar \Delta}_{{\rm h4}}/{\bar \Delta}_{{\rm h1}}>0$ indicates the absence of horizontal node. ${\bar \Delta}_{{\rm h4}}/{\bar \Delta}_{{\rm h1}}<0$ indicates the presence of horizontal node.
}
\label{fig:anisotropy}
\end{figure}

In previous subsection, 
we explained that the horizontal node is absent 
(${\bar \Delta}_{{\rm h4}}{\bar \Delta}_{{\rm h1}}>0$) 
due to inter-orbital attractive interaction. 
Here, we verify that this feature is very robust 
for wide parameter range.
Figure\,\ref{fig:anisotropy} is the impurity-dependence of the averaged gap size ratio ${\bar \Delta}_{{\rm h4}}/{\bar \Delta}_{\rm h1}$, where averaged gap size ${\bar \Delta}_{\alpha}$ is given as
\begin{eqnarray}
{\bar \Delta}_{\alpha}=\frac{\oint d{\bm k}_{\alpha}\Delta_{\alpha}({\bm k}_{\alpha})}{\oint d{\bm k}_{\alpha}}.
\end{eqnarray}
We discuss the results only for $\alpha_{s}=0.96$, 
since the results are almost independent of $\alpha_{s}$.
In the RPA analysis at $J/U=0.1$, we obtain ${\bar \Delta}_{{\rm h4}}/{\bar \Delta}_{{\rm h1}}\simeq -0.1$ 
in the whole range of $n_{{\rm imp}}$.
The result is almost independent of $J/U$. It indicates that horizontal node is robust in the RPA even if impurity effect is considered. 
In contrast, we obtain ${\bar \Delta}_{{\rm h4}}/{\bar \Delta}_{{\rm h1}}>0$ in the present theory with {\it U}-VCs, which means that the horizontal node is absent.
We find that the ratio ${\bar \Delta}_{{\rm h4}}/{\bar \Delta}_{{\rm h1}}$ becomes large with increasing $n_{{\rm imp}}$, from $0.5\, (n_{{\rm imp}}=0)$ to $0.8\,(n_{{\rm imp}}=0.05)$ at $J/U=0.1$. 
The horizontal node is absent even for $J/U=0.14$.
We stress that the horizontal node is absent even in the $s_{+-}$ wave state in the present theory.
The ratio ${\bar \Delta}_{{\rm h4}}/{\bar \Delta}_{{\rm h1}}$ is predicted to increase with $n_{{\rm imp}}$. 

\section {Summary}
\label{sec:summary}
In this paper, we studied the superconducting gap structure 
of BaFe$_2$(As,P)$_2$
based on the realistic two-dimensional five-orbital model.
The $U$-VCs due to the AL-processes are taken into account,
not only for the charge susceptibilities
but also for the superconducting gap equation.
Based on the present beyond-ME formalism, 
the nodal $s$-wave state or fully-gapped $s_{++}$-wave state is naturally obtained 
for wide parameter range.
The nodes appear only on the electron-FSs,
at which the orbital character changes between $xz(yz)$ and $xy$.
This result means the emergence of the loop-nodes on the electron-FSs,
consistently with the theoretical prediction in 
Ref. \cite{Saito-loopnode}.
In contrast, the gap functions on all hole-FSs are 
always fully-gapped, including the $z^2$-orbital outer hole-FS.
The obtained gap structure is consistent with the ARPES studies 
\cite{Shimojima,Yoshida}
and angle-resolved
thermal conductivity measurement \cite{Yamashita}.
The obtained nodal $s$-wave state changes to fully-gapped 
$s_{++}$-wave state
by introducing small amount of impurities,
accompanied by small reduction in $T_{\rm c}$.

The obtained results are essentially similar to the results 
of our previous RPA study \cite{Saito-loopnode},
in which a phenomenological inter-orbital 
quadrupole interaction term was introduced to
realize strong inter-orbital fluctuations in four $d$-orbitals.
In the present study, it is confirmed that the following 
nontrivial results are derived from the on-site Coulomb interaction
by considering the $U$-VCs,
without introducing any phenomenological interaction terms:
(i) Strong ferro-orbital and antiferro-orbital fluctuations
associated with four $d$-orbitals develop in Ba122 systems.
(ii)  Nearly isotropic gap function appears 
on all hole-FSs, including the $z^2$-orbital FS.
That is, the relation $\Delta_{h,z^2} \approx \Delta_{h,t_{2g}}$ holds.
(iii) Nodal gap structure appears on the electron-type FSs,
which corresponds to the loop-node discussed in Ref. \cite{Saito-loopnode},
for wide parameter range.
(iv) The obtained nodal $s$-wave state changes to the fully-gapped
$s_{++}$-wave state by introducing small amount of impurities,
accompanied by small reduction in $T_{\rm c}$.
These obtained results satisfactorily explain the 
characteristic superconducting gap structure observed in Ba122 pnictides.
The present gap equation beyond the standard ME formalism 
should be useful for understanding the 
rich variety of the superconducting states in various Fe-based superconductors,
such as La1111 \cite{Onari-PRL2014}, 
FeSe \cite{Yamakawa-PRB2017}, 
and Ba122 compounds.

\acknowledgements
We are grateful to Y. Matsuda, T Shibauchi, S. Shin, 
A. Fujimori, T. Shimojima, T. Yoshida, K. Okazaki, and S. Onari
for fruitful discussions.
This work was supported by Grant-in-Aid for Scientific Research from 
the Ministry of Education, Culture, Sports, Science, and Technology, Japan.

\appendix

\section{The reason why spin fluctuation mediate attractive interaction}
\label{sec:appendix}
\begin{figure}[htbp]
\includegraphics[width=1.0\linewidth]{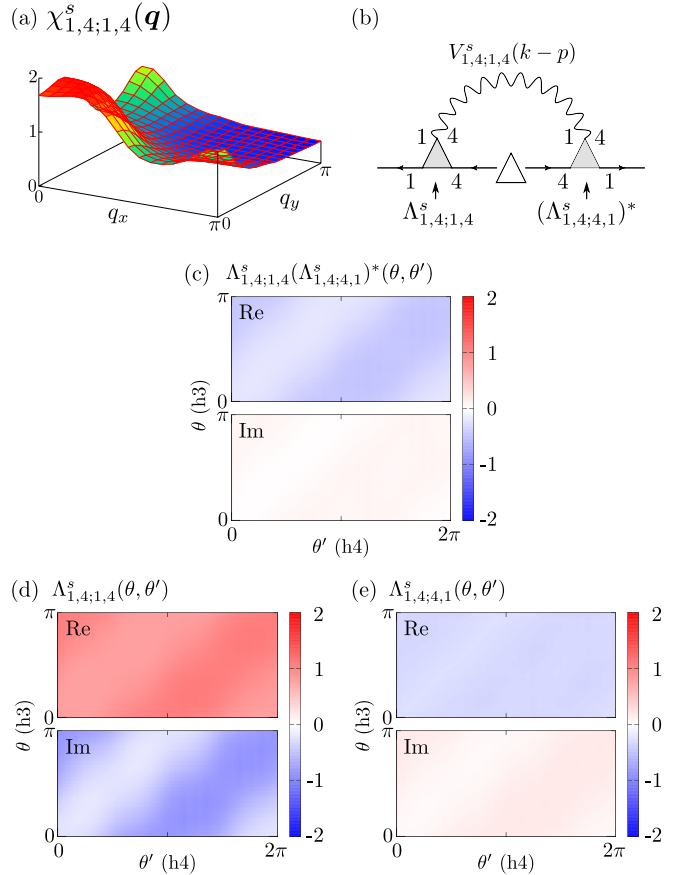}
\caption{The spin susceptibility $\chi^{s}_{1,4;1,4}({\bm q})$. (b) The expression of interaction given by $\chi^{s}_{1,4;1,4}({\bm q})$. $V^{s}_{1,4;1,4}({\bm q})$ is multiplied by $\Lambda^{s}_{1,4;1,4}(\Lambda^{s}_{1,4;4,1})^{\ast}$. (c) The real and imaginary parts of $\Lambda^{s}_{1,4;1,4}(\Lambda^{s}_{1,4;4,1})^{\ast}$. Red and blue colors respectively means positive and negative interactions.
Re$[\Lambda^{s}_{1,4;1,4}(\Lambda^{s}_{1,4;4,1})^{\ast}]<0$ means that off-diagonal spin channel {\it U}-VCs change the sign of interaction given by $\chi^{s}_{1,4;1,4}({\bm q})$. The real and imaginary parts of (d) $\Lambda^{s}_{1,4;1,4}$ and (e) $\Lambda^{s}_{1,4;4,1}$.
}
\label{fig:Lambda}
\end{figure}
In the main text, we showed the averaged $V^{(1)s}_{\Lambda}$ in Fig.\,\ref{fig:vpair}(c). 
Interestingly, ${\bar V}^{(1)s}_{\Lambda}(\alpha,\beta)$ for $(\alpha,\beta)=({\rm h1,h3}),({\rm h2,h3})$ and $({\rm h3,h4})$ are attractive,
although their contribution to the pairing is very small.
This nontrivial result originates from inter-orbital 
{\it U}-VCs for spin channel, since 
spin fluctuations always give repulsive interaction
in the ME formalism.
We explain the reason why spin fluctuations cause attractive interaction by focusing on ${\bar V}^{(1)s}_{\Lambda}({\rm h3},{\rm h4})$.

Both $\chi^{s}_{1,1;1,1}({\bm q})$ and $\chi^{s}_{1,4;4,1}({\bm q})$ give repulsive interaction even if we consider the spin channel {\it U}-VC, because $V^{s}_{1,1;1,1}({\bm q})$ and $V^{s}_{1,4;4,1}({\bm q})$ are multiplied by $|\Lambda^{s}_{1,1;1,1}|^{2}>0$ and $|\Lambda^{s}_{1,4;1,4}|^{2}>0$, respectively.
The sign is not changed, and therefore these spin fluctuations give repulsive interaction.
In contrast, the sign of interaction caused by $\chi^{s}_{1,4;1,4}({\bm q})$ (shown in Fig.\,\ref{fig:Lambda}(a)) is nontrivial.
Fig.\,\ref{fig:Lambda}(b) is the expression of the interaction given by $\chi^{s}_{1,4;1,4}({\bm q})$. 
$V^{s}_{1,4;1,4}({\bm q})$ is multiplied by $\Lambda^{s}_{1,4;1,4}(\Lambda^{s}_{1,4;4,1})^{\ast}$, of which the sign is nontrivial. 
In the present model, the real part of $\Lambda^{s}_{1,4;1,4}(\Lambda^{s}_{1,4;4,1})^{\ast}$ is negative as shown in the upper panel of Fig.\,\ref{fig:Lambda}(c). 
Due to off-diagonal components of spin channel {\it U}-VC, the sign of interaction given by $\chi^{s}_{1,4;1,4}({\bm q})$ is changed.
The imaginary part of $\Lambda^{s}_{1,4;1,4}(\Lambda^{s}_{1,4;4,1})^{\ast}$ shown in the lower panel in Fig.\,\ref{fig:Lambda}(c) is canceled out by Im$[\Lambda^{s}_{1,4;4,1}(\Lambda^{s}_{1,4;1,4})^{\ast}]$. 
We note that Re$[\Lambda^{c}_{1,4;1,4}(\Lambda^{c}_{1,4;4,1})^{\ast}]$ is positive, thus interaction given by inter-orbital charge susceptibility $\chi^{c}_{1,4;1,4}({\bm q})$ is attractive.
In Figs.\,\ref{fig:Lambda}(d) and (e), we show the real and imaginary parts of $\Lambda^{s}_{1,4;1,4}$ and $\Lambda^{s}_{1,4;4,1}$. 
Re$[\Lambda^{s}_{1,4;1,4}(\Lambda^{s}_{1,4;4,1})^{\ast}]$ is given as ${\rm Re}[\Lambda^{s}_{1,4;1,4}]{\rm Re}[\Lambda^{s}_{1,4;4,1}]+{\rm Im}[\Lambda^{s}_{1,4;1,4}]{\rm Im}[\Lambda^{s}_{1,4;4,1}]$. 
The main contribution originates from ${\rm Re}[\Lambda^{s}_{1,4;1,4}]{\rm Re}[\Lambda^{s}_{1,4;4,1}]$, because $|{\rm Im}[\Lambda^{s}_{1,4;4,1}]|\ll 1$.
For this reason,  pairing interaction caused by $\chi^{s}_{1,4;1,4}({\bm q})$ is attractive due to the off-diagonal components of spin channel {\it U}-VC. The other attractive panels in Fig.\,\ref{fig:vpair}(c) are understood in the same way.



\end{document}